\documentclass{elsart}
\usepackage{amsfonts}
\usepackage{amsmath}

\input{tcilatex}

\begin{document}

\begin{frontmatter}%

\title{The number and probability of canalizing functions}%

\author{Winfried Just}%

\address{corresponding author; just@math.ohiou.edu, phone: (740)-593-1260, fax: (740)-593-9805, Department of Mathematics, Ohio University, Athens, Ohio 45701, U.S.A.}%

\collab{Ilya Shmulevich}%

\address{is@ieee.org, Cancer Genomics Laboratory, University of Texas M. D. Anderson Cancer Center, Houston, Texas 77030, U.S.A.}%

\collab{John Konvalina}%

\address{johnkon@unomaha.edu, Department of Mathematics University of Nebraska at Omaha Omaha, NE 68182-0243, U.S.A.}%

\begin{abstract}
Canalizing functions have important applications in physics and biology. For
example, they represent a mechanism capable of stabilizing chaotic behavior
in Boolean network models of discrete dynamical systems. When comparing the
class of canalizing functions to other classes of functions with respect to their
evolutionary plausibility as emergent control rules in genetic regulatory
systems, it is informative to know the number of canalizing functions with a
given number of input variables. This is also important in the context of
using the class of canalizing functions as a constraint during the inference
of genetic networks from gene expression data. To this end, we derive an
exact formula for the number of canalizing Boolean functions of $n$
variables. We also derive a formula for the probability that a random
Boolean function is canalizing for any given bias $p$ of taking the value 1.
In addition, we consider the number and probability of Boolean functions
that are canalizing for exactly $k$ variables. Finally, we provide an
algorithm for randomly generating canalizing functions with a given bias $p$
and any number of variables, which is needed for Monte Carlo simulations of
Boolean networks.%
\end{abstract}%

\begin{keyword}%
canalizing function, forcing function, Boolean network%
\end{keyword}%
\thanks{PACS codes: 02.10.Eb, 05.45.+b, 87.10.+e}

\end{frontmatter}%

\section{Introduction}

A Boolean function (on $n$ variables) is a function $f:\{0,1\}^{n}%
\rightarrow \{0,1\}$. A canalizing function (also called a forcing function)
is a type of Boolean function in which at least one of the input variables
is able to determine the function output regardless of the values of the
other variables. For example, the function $f\left( x_{1},x_{2},x_{3}\right)
=x_{1}+x_{2}x_{3}$, where the addition symbol stands for disjunction and the
multiplication for conjunction, is a canalizing function, since setting $%
x_{1}$ to 1 guarantees that the function value is 1 regardless of the value
of $x_{2}$ or $x_{3}$. On the other hand, the function $f\left(
x_{1},x_{2}\right) =x_{1}\oplus x_{2}$, where $\oplus $ is addition modulo
2, is not a canalizing function, since the values of both variables always
need to be known in order to determine the function output.

Canalizing functions have been implicated in a number of phenomena related
to discrete dynamical systems as well as nonlinear filters. Concerning the
latter, they have been used to study the convergence behavior of an
important class of nonlinear digital filters called stack filters \cite%
{wendt,gabbouj,yu}. For example, stack filters defined by canalizing
functions are known to possess a convergence property whereby a filter is
guaranteed to converge to a so-called root signal or fixed point of the
filter after a finite number of passes \cite{gabbouj}. In \cite{yucoyle},
some learning schemes were proposed to find minimal filters defined by
canalizing functions.

Canalizing functions also play an important role in the study of phase
transitions in random Boolean networks \cite%
{kauffmanbook,kauffman90,lynch,stauffer,ilyapost}. Boolean networks have
been one of the most intensively studied models of discrete dynamical
systems and have been used to gain insight into the behavior of large
genetic networks \cite{kauffmanbook}, evolutionary principles \cite%
{stern,bornholdt}, and the development of chaos \cite{glass,bhat}. Although
structurally simple, these systems are capable of displaying a remarkably
rich variety of complex behavior. Canalizing functions represent one of the
few known mechanisms capable of preventing chaotic behavior in Boolean
networks \cite{kauffmanbook}. By increasing the percentage of canalizing
functions in a Boolean network, one can move closer toward the ordered
regime and, depending on the connectivity and the distribution of the number
of canalizing variables, cross the phase transition boundary \cite%
{kauffmaninvestigations}. In fact, there is overwhelming evidence that
canalizing functions are abundantly utilized in higher vertebrate gene
regulatory systems~\cite{kauffmanbook}. A recent large-scale study of the
literature on transcriptional regulation in eukaryotes demonstrated an
overwhelming bias towards canalizing rules \cite{harris}. Canalization is
also a natural mechanism for designing robustness against noise \cite%
{sawhill}.

Knowledge of the number of possible canalizing functions with a given number
of input variables is important for determining the degree to which these
functions are evolutionarily plausible as regulatory rules in genetic
networks. There are two related issues here. First, a class of functions
that is overly limited in size is unlikely to emerge via the mechanism of
random selection. Thus, when comparing different classes of functions vis-%
\`{a}-vis their likelihood of giving rise to regulatory control rules, it is
informative to know their respective sizes \cite{ilyapost}. Second, when
gene regulatory rules are inferred from real gene expression measurements 
\cite{harri}, it is often beneficial to constrain the inferential algorithms
to a certain class of functions that can be produced. It may seem that
imposing a constraint (e.g., restricting all functions to be canalizing) can
only result in a degradation of the performance of the algorithm, thus
yielding a larger estimation or prediction error relative to an algorithm
with no imposed constraints. But it turns out that doing so can often
improve the tractability and precision of the inference. This can be
particularly noticeable when an inference is made from small sample sizes.
In order to quantify the reduction in `design cost' owing to the constraint,
it is again informative to consider the size of the class of functions used
as a constraint. Thus, it is an important goal to establish the number of
canalizing functions of a given number of input variables.

Of course, one approach is to generate all Boolean functions with $n$
variables and check whether each one is canalizing. However, despite
efficient methods to test the canalizing property \cite{spectral}, this
approach becomes prohibitive for large values of $n$ and the exact number
has only been known for $n\leq 5$ \cite{ilyapost}. It has also been known
that the number of canalizing functions with $n$ variables is upper bounded
by $4n\cdot 2^{2^{n-1}}$ \cite{aldana}.

In this paper, we derive an exact formula for the number of canalizing
functions with $n$ variables. In addition, we also derive a formula for the
probability that a random Boolean function whose truth table is a Bernoulli$%
\left( p\right) $ random vector is canalizing. The latter is important
because the `bias' $p$ of Boolean functions also plays a crucial role in the
order-disorder transition in Boolean networks and it is known that
canalizing functions are likely to be biased, meaning that they are expected
to have a large number of ones or zeros in their truth tables \cite%
{stauffer,ilyapost}. Since a canalizing function can have one or more
canalizing variables, we also consider the number and probability of Boolean
functions that are canalizing for exactly $k$ variables. This is also a
relevant issue because it is known that tuning the number of canalizing
inputs in a random Boolean network can dramatically affect its dynamical
behavior. Moreover, according to the formulas derived in our paper, real
genetic regulatory rules appear to be highly skewed towards large numbers of
canalizing inputs \cite{harris} relative to what should be expected by
chance in a canalizing function.

\section{The probability of canalizing functions}

Throughout this paper, let $n\geq 1$ be a fixed positive integer. For each
positive integer $k$, the set $\{0,1,\ldots ,k-1\}$ will be denoted by $[k]$%
. The cardinality of a set $A$ will be denoted by $|A|$. Let $0\leq p\leq 1$%
. We will consider the following probability measure on the space of all
Boolean functions: 
\begin{equation*}
Pr_{p}(f)=p^{|f^{-1}\{1\}|}(1-p)^{|f^{-1}\{0\}|}.
\end{equation*}%
We call $Pr_{p}(f)$ the probability of $f$ for bias $p$.

Recall that a Boolean function $f$ is canalizing if there exist $i \in n$
(called a canalizing variable) and $s, v \in \{ 0, 1\}$ such that:

\begin{equation}
\forall x \in \{0,1\}^n\,(x_i = s\Rightarrow f(x_i)=v).  \label{canaldef}
\end{equation}%
If $v=1$, then we will say that $f $ is positively canalizing; if $v=0$,
then we will say that $f$ is negatively canalizing.

Let $C$ be the set of all canalizing Boolean functions; let $PC$ be the set
of all positively canalizing Boolean functions, let $NC$ be the set of all
negatively canalizing Boolean functions, and let $BC$ be the set of Boolean
functions that are both positively and negatively canalizing.

Our goal in this section is to calculate $Pr_p (C)$. It is clear that

\begin{equation}  \label{PC+NC}
Pr_p (C) = Pr_p (PC) + Pr_p (NC) - Pr_p(BC).
\end{equation}

Let us first dispose of the easy part and calculate $Pr_{p}(BC)$. Note that
it cannot be the case that a Boolean function $f $ is positively canalizing
for a canalizing variable $x_i$ and negatively canalizing for a different
canalizing variable $x_j\neq x_i$. Thus for every $f\in BC$ there exists a
unique canalizing variable $x_i(f)$, and we either have

\begin{equation*}
\forall x \in \{ 0,1\}^n (x_i = 0 \Rightarrow f(x_i) = 0) \ \& (x_i = 1
\Rightarrow f(x_i) = 1),
\end{equation*}

or we have 
\begin{equation*}
\forall x \in \{ 0,1\}^n (x_i = 0 \Rightarrow f(x_i) = 1) \ \& (x_i = 1
\Rightarrow f(x_i) = 0).
\end{equation*}

Thus $|BC| = 2n$; and for every function $f \in BC$ we have $Pr_p (f) =
p^{2^{n-1}}(1-p)^{2^{n-1}}$. It follows that

\begin{equation}  \label{PrBC}
Pr_p (BC) = 2n p^{2^{n-1}}(1-p)^{2^{n-1}}.
\end{equation}

In our calculations of $Pr_p (PC)$ and $Pr_p (NC)$ it will be convenient to
work only with nonconstant canalizing functions. Let $PC^- = PC \backslash
\{ \mathbf{1}\}$ and $NC^- = NC \backslash \{ \mathbf{0}\}$, where $\mathbf{1%
}, \mathbf{0}$ are the Boolean functions that take the value $1$
respectively $0$ everywhere. In this terminology, equation~(\ref{PC+NC}) is
equivalent to: 
\begin{equation}  \label{PC-NC-}
Pr_p (C) = Pr_p (PC^-) + Pr_p (NC^-) + p^{2^n} + (1-p)^{2^n} - 2n
p^{2^{n-1}}(1-p)^{2^{n-1}}.
\end{equation}

Now we need to be a little more specific about the number of variables for
which a function is canalizing.

\begin{definition}
\label{Icandef} Let $f: \{ 0, 1 \}^n \rightarrow \{ 0 , 1 \}$ and let $I$ be
a nonempty subset of $[n]$. We say that $f$ is positively canalizing on $I$
if there exists a function $\sigma : I \rightarrow \{ 0, 1\}$ called a
signature of $f$ on $I$ such that

\begin{equation}  \label{signature}
\forall x \in \{ 0, 1 \}^n \ ((\exists i \in I \, x_i \neq \sigma (i))
\Rightarrow f(x_i) = 1).
\end{equation}

The notion of being negatively canalizing on $I$ is defined analogously. The
set of all nonconstant Boolean functions that are positively canalizing on a
given index set $I$ will be denoted by $PC^-_I$; the set of all nonconstant
Boolean functions that are negatively canalizing on a given index set $I$
will be denoted by $NC^-_I$.
\end{definition}

\begin{fact}
\label{uniquesignature} Let $f \in PC^-_I$ or $f \in NC^-_I$. Then there
exists exactly one signature for $f$ on $I$.
\end{fact}

\begin{proof}
Without loss of generality suppose $f \in PC^-_I$, and assume towards a contradiction that $\sigma, \tau : I
\rightarrow \{ 0, 1\}$ are two different signatures for $f$.  Let $i \in I$ be such that 
$\sigma (i) \neq \tau (i)$.  Then for every $x \in \{ 0, 1\}^n$ we have $x_i \neq \sigma (i)$
or $x_i \neq \tau (i)$, and it follows from equation~(\ref{signature}) that $f(x) = 1$.
Thus $f = \mathbf{1}$, which contradicts the assumption that $f \in PC^-_I$.
\end{proof}

If $f \in PC^-_I$ or $f \in NC^-_I$, then we let $\sigma_{I, f}$ denote the
unique signature of $f$ on $I$.

\begin{lemma}
\label{intersection} Let $I$ be a nonempty subset of $[n]$. Then $PC^-_I =
\bigcap_{i \in I} PC^-_{\{ i\} }$ and $NC^-_I = \bigcap_{i \in I} NC^-_{\{
i\} }$.
\end{lemma}

\begin{proof}
Suppose $f  \in PC^-_I$ and $i \in I$.  It is easy to see that the restriction of $\sigma_{I, f}$ to $\{ i \}$ 
is a signature for $f$ on $\{ i \}$, and thus $f \in PC^-_{\{ i \}}$.  
Now suppose $f \in PC^-_{\{ i \} }$ for all $i \in I$.  Let 
$\sigma = \bigcup \{ \sigma_{\{ i \} , f}: \, i \in I \}$.  Then $\sigma$ is a signature for $f$ on $I$, and it follows that $f \in PC^-_I$. 

The proof of the second equation is analogous.
\end{proof}

It follows from the definition of canalizing functions that

\begin{equation}  \label{irepresentation}
PC^- = \bigcup_{i < n} PC^-_{\{ i \} } , \qquad \qquad NC^- = \bigcup_{i <
n} NC^-_{\{ i \} }
\end{equation}

Unfortunately, the sets $PC^-_{\{ i \} }$ are not pairwise disjoint. So we
have to use the Inclusion-Exclusion Principle to calculate the probability
of the union of these sets. This gives:

\begin{equation}  \label{prsum1}
\begin{aligned} &Pr_p (PC^-) = \sum_{0 \leq i < n} Pr_p (PC^-_{ \{ i \} }) -
\sum_{0 \leq i_1 < i_2 < n} Pr_p(PC^-_{\{ i_1\} } \cap PC^-_{\{ i_2\} }) +
\dots \\ &+ (-1)^{k+1} \sum_{0 \leq i_1 < i_2 < \dots < i_k < n} Pr_p
(PC^-_{\{ i_1 \} } \cap PC^-_{\{ i_2 \} } \cap \dots \cap PC^-_{\{ i_k \} })
+ \dots . \end{aligned}
\end{equation}

By Lemma~\ref{intersection}, equation~(\ref{prsum1}) can be written as:

\begin{equation}  \label{prsum2}
Pr_p (PC^-) = \sum_{k=1}^n (-1)^{k+1} \sum_{0 \leq i_1 < i_2 < \dots < i_k <
n} Pr_p (PC^-_{\{ i_1 ,i_2, \dots , i_k \} }).
\end{equation}

Since for $|I| = |J|$ we obviously have $Pr_p (PC^-_I) = Pr_p (PC^-_J)$, we
can rewrite equation~(\ref{prsum2}) as follows:

\begin{equation}  \label{prsum3}
Pr_p (PC^-) = \sum_{k=1}^n (-1)^{k+1} \binom{n}{k} Pr_p (PC^-_{[k]}).
\end{equation}

The analogous reasoning shows that

\begin{equation}  \label{prsum4}
Pr_p (NC^-) = \sum_{k=1}^n (-1)^{k+1} \binom{n}{k} Pr_p (NC^-_{[k]}).
\end{equation}

Now it remains to compute $Pr_p(PC^-_{[k]})$ and $Pr_p(NC^-_{[k]})$.

\begin{lemma}
\label{PCklemma} Let $1 \leq k < n$. Then 
\begin{equation}  \label{PCk}
Pr_p(PC^-_{[k]}) = 2^k(p^{2^n - 2^{n-k}} - p^{2^n}).
\end{equation}
\begin{equation}  \label{NCk}
Pr_p(NC^-_{[k]}) = 2^k((1-p)^{2^n - 2^{n-k}} - (1-p)^{2^n}).
\end{equation}
\end{lemma}

\begin{proof} We prove equation~(\ref{PCk}); the proof of equation~(\ref{NCk}) is analogous.
By Fact~\ref{uniquesignature} we have 

\begin{equation}\label{PCk1}
Pr_p(PC^-_{[k]}) = \sum_{\sigma \in \{ 0, 1\}^n} Pr_p (f \in PC^-_{[k]} \ \& \ \sigma_{[k], f} = \sigma ).
\end{equation}

It is clear that for any $\sigma , \tau: [k] \rightarrow \{ 0, 1\}$
we have $Pr_p (f \in PC^-_{[k]} \ \& \ \sigma_{[k], f} = \sigma ) =
Pr_p (f \in PC^-_{[k]} \ \& \ \sigma_{[k], f} = \tau )$.
Pick an arbitrary $\sigma^*: [k] \rightarrow \{ 0, 1\}$.  Equation~(\ref{PCk1}) now implies:

\begin{equation}\label{PCk2}
Pr_p(PC^-_{[k]}) = 2^k Pr_p (f \in PC^-_{[k]} \ \& \ \sigma_{[k], f} = \sigma^* ).
\end{equation}

Let us calculate $Pr_p (f \in PC^-_{[k]} \ \& \ \sigma_{[k], f} = \sigma^* )$. 
If $f \in PC^-_{[k]}$ and $\sigma_{[k], f} = \sigma^*$, then $f (x) = 1$ whenever the restriction of
$x$ to the first $k$ variables is not equal to 
$\sigma^*$.  So there are $2^{n-k}$ arguments $x$ of $f$ on which $f$ can take
arbitrary values (except taking value $1$ everywhere), 
and $2^n - 2^{n-k}$ arguments $x$ where $f$ has to take value $1$.  In other words,

\begin{equation}\label{disjunction}
Pr_p ((f \in PC^-_{[k]} \ \& \ \sigma_{[k], f} = \sigma^* ) \ \vee f = \mathbf{1}) = p^{2^n - 2^{n-k}}.
\end{equation}

Since $Pr_p(\mathbf{1}) = p^{2^n}$, equation~(\ref{disjunction}) implies 

\begin{equation}\label{PCk*}
Pr_p (f \in PC^-_{[k]} \ \& \ \sigma_{[k], f} = \sigma^*) = p^{2^n - 2^{n-k}} - p^{2^n}.
\end{equation}

This in turn implies equation~(\ref{PCk}).
\end{proof}

Now let us put all our formulas together. We get:

\begin{equation}  \label{semifinal}
\begin{aligned} &Pr_p (C) = p^{2^n} + (1-p)^{2^n} - 2n
p^{2^{n-1}}(1-p)^{2^{n-1}} + \\ &\sum_{k=1}^n (-1)^{k+1} \binom{n}{k} 2^k
(p^{2^n - 2^{n-k}} + (1-p)^{2^n - 2^{n-k}} - p^{2^n} - (1-p)^{2^n}).
\end{aligned}
\end{equation}

Note that

\begin{equation}  \label{simplifier}
\begin{aligned} &\sum_{k=1}^n (-1)^{k+1} \binom{n}{k} 2^k (- p^{2^n} -
(1-p)^{2^n}) = \\ &(p^{2^n} + (1-p)^{2^n})\sum_{k=1}^n \binom{n}{k} (-2)^k =
\\ &(p^{2^n} + (1-p)^{2^n})((1-2)^n - 1) = (p^{2^n} + (1-p)^{2^n})((-1)^n -
1). \end{aligned}
\end{equation}

Thus equation~(\ref{semifinal}) simplifies to:

\begin{equation}  \label{final}
\begin{aligned} Pr_p(C) =& \  (-1)^n (p^{2^n} + (1-p)^{2^n}) - 2n
p^{2^{n-1}}(1-p)^{2^{n-1}} + \\ &\sum_{k=1}^n (-1)^{k+1} \binom{n}{k} 2^k
(p^{2^n - 2^{n-k}} + (1-p)^{2^n - 2^{n-k}}). \end{aligned}
\end{equation}

\section{The number of canalizing functions}

Equation~(\ref{final}) allows us to derive a formula for the number of
canalizing functions as follows. Set $p = 0.5$. Then all functions have
equal probability, and we simply can compute:

\begin{equation}
\begin{aligned} &|C| = Pr_{0.5} (C) 2^{2^n} = \\ &2^{2^n}2(((-1)^n -
n)2^{-2^n} + \sum_{k=1}^n (-1)^{k+1} \binom{n}{k} 2^k 2^{-2^n + 2^{n-k}}) =
\\ &2((-1)^n - n) + \sum_{k=1}^n (-1)^{k+1} \binom{n}{k} 2^{k +
1}2^{2^{n-k}}. \end{aligned}  \label{number}
\end{equation}%
The values of $|C|$ for $n=1,\ldots ,10$ are shown in Table \ref{TableKey}.

It is interesting to note that for large $n$ the value of $|C|$ given by
equation~(\ref{number}) asymptotically approaches the upper bound of $%
4n\cdot 2^{2^{n-1}}$ given in~\cite{aldana}. To see this, let $S_k = \binom{n%
}{k}2^{k+1}2^{2^{n-k}}$ for $1 \leq k \leq n$. Then

\begin{equation}  \label{number1}
|C| = 2((-1)^n - n)+ \sum_{k=1}^n (-1)^{k+1} S_k.
\end{equation}

For sufficiently large $n$, the first term becomes negligible, and we can
concentrate on the asymptotic behavior of $S = \sum_{k=1}^n (-1)^{k+1} S_k$.
Moreover, it is not hard to see that $S_k > S_{k+1}$ for all $k< n$. Thus
the partial sums of $S$ with an odd number of terms form an upper bound for $%
S$, while the partial sums with an even number of terms form a lower bound.
In particular, for the first and second partial sums we have the
inequalities:

\begin{equation}  \label{parsum1}
S_1 - S_2 \leq S \leq S_1.
\end{equation}

Dividing by $S_1$ we obtain

\begin{equation}  \label{parsum2}
1 - \frac{S_2}{S_1} \leq \frac{S}{S_1} \leq 1.
\end{equation}

As $n$ approaches infinity, $\frac{S_2}{S_1}$ approaches zero, and
therefore, $S$ is asymptotic to $S_1$. Now it suffices to note that $S_1 = 
\binom{n}{1}2^{1+1}2^{2^{n-1}} = 4n\cdot 2^{2^{n-1}}$ is exactly the upper
bound given in~\cite{aldana}.

\section{Functions that are canalizing for $k$ variables}

By definition, a Boolean function is canalizing if and only if it is
canalizing for at least one variable. How can we compute the number and
probability of Boolean functions that are canalizing for exactly $k$
variables? To solve this problem, we need a generalization of the
Inclusion-Exclusion Principle. The following lemma appears as Corollary 5B.4
in \cite{williamson}.

\begin{lemma}
\label{genincex} Let $f_0, \ldots f_{n-1}$ be real-valued functions with a
common domain, and let $u$ be the function that is identically $1$ on the
common domain of the $f_i$'s. Let $I \subseteq [n]$, and let $I^c = [n]
\backslash I$. Then 
\begin{equation}  \label{williamson}
\prod_{i\in I}f_i\prod_{i \in I^c} (u - f_i) = \sum_{R \supseteq I}
(-1)^{|R|-|I|}\prod_{i \in R} f_i.
\end{equation}
\end{lemma}

Now suppose that $E_0, \ldots , E_{n-1}$ are events in a fixed probability
space $\Omega$, and that $f_i$ is the characteristic function of $E_i$ on $%
\Omega$ for $i = 0, \ldots , n-1$. Then we have for $I \subseteq [n]$: 
\begin{equation*}
Pr (\prod_{i\in I} f_i = 1) = Pr(\bigcap_{i \in I} E_i),
\end{equation*}
\begin{equation*}
Pr (\prod_{i\in I} f_i \prod_{i \in I^c} (u - f_i) = 1) = Pr(\bigcap_{i \in I}
E_i \cap \bigcap_{i \in I^c} E_i^c),
\end{equation*}
and equation~(\ref{williamson}) translates into:

\begin{equation}  \label{prwilliamson}
Pr(\bigcap_{i \in I} E_i \cap \bigcap_{i \in I^c} E_i^c) = \sum_{R \supseteq
I} (-1)^{|R|-|I|} Pr(\bigcap_{i \in R} E_i).
\end{equation}

Now let $E_0, \ldots , E_{n-1}$ and $I$ be as above. Following
Definition~5B.5 of~\cite{williamson} we define:

\begin{equation*}
IN(I) = \{ \omega \in \Omega: \, \omega \in E_i \Leftrightarrow i \in I\},
\end{equation*}
and for $k \leq n$: 
\begin{equation*}
IN(k) = \bigcup_{|I|=k} IN(I).
\end{equation*}

The following lemma is a straightforward generalization of Corollary~5B.6 of~%
\cite{williamson}:

\begin{lemma}
\label{williamsonlemma} In the terminology introduced above we have:

\begin{equation}  \label{wl1}
Pr(IN(I)) = \sum_{R \supseteq I} (-1)^{|R|-|I|} Pr(\bigcap_{i \in R} E_i).
\end{equation}

\begin{equation}  \label{wl2}
Pr((IN(k)) = \sum_{r=k}^n \binom{r}{k} (-1)^{r-k} \sum_{R \subseteq [n], |R|
= r} Pr(\bigcap_{i \in R} E_i).
\end{equation}
\end{lemma}

Let us apply equation~(\ref{wl2}) to the situation where $\Omega$ is the
space of all Boolean functions of $n$ variables with probability function $%
Pr_p$ defined above. For $1 \leq k \leq n$, let $Pr_p(PCE_k)$ denote the
probability that a randomly chosen Boolean function with bias $p$ is
positively (but not negatively) canalizing on exactly $k$ variables, and let 
$Pr_p(NCE_k)$ denote the probability that a randomly chosen Boolean function
with bias $p$ is negatively (but not positively) canalizing on exactly $k$
variables. We will compute $Pr_p(PCE_k)$. For $i < n$, let $E_i =
PC_{\{i\}}^-$. Note that for this choice of $E_i$ and $k < n$, $IN(k)$ is
the set of functions that are positively canalizing for exactly $k$
variables; and the set of functions that are positively canalizing for $n$
variables is $IN(n) \cup \{ \mathbf{1}\}$. In view of Lemma~\ref%
{intersection}, equation~(\ref{wl2}) now boils down to the following:

\begin{equation}  \label{wl3}
Pr_p((IN(k)) = \sum_{r=k}^n \binom{r}{k} (-1)^{r-k} \sum_{R \subseteq [n],
|R| = r} Pr_p(PC_R^-).
\end{equation}

Since the probability of $PC_R^-$ depends only on $|R|$, we get

\begin{equation}  \label{wl4}
Pr_p((IN(k)) = \sum_{r=k}^n \binom{r}{k} (-1)^{r-k} \binom{n}{r}
Pr_p(PC_{[r]}^-).
\end{equation}

It follows from Lemma~\ref{PCklemma} that for $k > 1$ we have:

\begin{equation}  \label{wl5}
\begin{aligned} &Pr_p((IN(k)) =\sum_{r=k}^n \binom{r}{k} (-1)^{r-k}
\binom{n}{r} Pr_p(PC_{[r]}^-) = \\ &\sum_{r=k}^n \binom{r}{k} (-1)^{r-k}
\binom{n}{r} 2^r (p^{2^n - 2^{n-r}} - p^{2^n} ). \end{aligned}
\end{equation}

Thus it follows that for $1 < k < n$ we have:

\begin{equation}  \label{wl5a}
Pr_p(PCE_k) = \sum_{r=k}^n \binom{r}{k} (-1)^{r-k} \binom{n}{r} 2^r (p^{2^n
- 2^{n-r}} - p^{2^n} ).
\end{equation}

A similar argument shows that for $1 < k < n$ we have:

\begin{equation}  \label{wl5b}
Pr_p(NCE_k) = \sum_{r=k}^n \binom{r}{k} (-1)^{r-k} \binom{n}{r} 2^r
((1-p)^{2^n - 2^{n-r}} - (1-p)^{2^n} ).
\end{equation}

For $1 < k = n$ we need to add the two constant functions:

\begin{equation}  \label{wl5c}
Pr_p(PCE_n) = 2^n(p^{2^n - 1} - p^{2^n}) + p^{2^n}.
\end{equation}

\begin{equation}  \label{wl5d}
Pr_p(NCE_n) = 2^n((1-p)^{2^n - 1} - p^{2^n}) + (1-p)^{2^n}.
\end{equation}

For $1 = k < n$ we need to subtract the probability that the function is
canalizing both ways. This gives:

\begin{equation}  \label{wl6}
\begin{aligned} &Pr_p(PCE_1) = n (Pr_p(PC_{[1]}^-) -
2p^{2^{n-1}}(1-p)^{2^{n-1}}) + \\ &\sum_{r=2}^n r (-1)^{r-1} \binom{n}{r}
(Pr_p(PC_{[r]}^-)) = \\ & 2n(p^{2^n - 2^{n - 1} } - p^{2^n} -
p^{2^{n-1}}(1-p)^{2^{n-1}}) + \\ &\sum_{r=2}^n r (-1)^{r-1} \binom{n}{r} 2^r
(p^{2^n - 2^{n - r} } - p^{2^n}) =\\ & 2n(p^{2^{n - 1} } - p^{2^n} -
p^{2^{n-1}}(1-p)^{2^{n-1}}) + \\ &\sum_{r=2}^n r (-1)^{r-1} \binom{n}{r} 2^r
(p^{2^n - 2^{n - r} } - p^{2^n} ). \end{aligned}
\end{equation}

\begin{equation}  \label{wl6n}
\begin{aligned} &Pr_p(NCE_1) = n (Pr_p(NC_{[1]}^-) -
2p^{2^{n-1}}(1-p)^{2^{n-1}}) + \\ &\sum_{r=2}^n r (-1)^{r-1} \binom{n}{r}
(Pr_p(NC_{[r]}^-)) = \\ & 2n((1-p)^{2^n - 2^{n - 1} } - (1-p)^{2^n} -
p^{2^{n-1}}(1-p)^{2^{n-1}}) + \\ &\sum_{r=2}^n r (-1)^{r-1} \binom{n}{r} 2^r
((1-p)^{2^n - 2^{n - r} } - (1-p)^{2^n}) =\\ & 2n((1-p)^{2^{n - 1} } -
(1-p)^{2^n} - p^{2^{n-1}}(1-p)^{2^{n-1}}) + \\ &\sum_{r=2}^n r (-1)^{r-1}
\binom{n}{r} 2^r ((1-p)^{2^n - 2^{n - r} } - (1-p)^{2^n} ). \end{aligned}
\end{equation}

Let $c(k)$ denote the number of functions that are canalizing for exactly $k$
variables. We have:

\begin{equation}  \label{Pck<n}
c(k) = (Pr_{0.5}(PCE_k) + Pr_{0.5}(NCE_k))2^{2^n} \qquad \mbox{if\ } 1 < k <
n,
\end{equation}

\begin{equation}  \label{Pck=n}
c(k) = 2 + (Pr_{0.5}(PCE_k) + Pr_{0.5}(NCE_k))2^{2^n} \qquad \mbox{if\ } 1 <
k = n,
\end{equation}

\begin{equation}  \label{P1=k<n}
c(k) = (Pr_{0.5}(PCE_1) + Pr_{0.5}(NCE_1) + Pr_{0.5}(BC))2^{2^n} \qquad %
\mbox{if\ } 1 = k < n,
\end{equation}

\begin{equation}  \label{P1=k=n}
c(k) = 2 + (Pr_{0.5}(PCE_1) + Pr_{0.5}(NCE_1) + Pr_{0.5}(BC))2^{2^n} \qquad %
\mbox{if\ } 1 = k = n.
\end{equation}

This implies the following formulas for $c(k)$:

For $1 < k < n$:

\begin{equation}  \label{1<k<n}
c(k) = \sum_{r=k}^n \binom{r}{k} (-1)^{r-k} \binom{n}{r} 2^{r+1}
(2^{2^{n-r}} - 1).
\end{equation}

For $1 < k = n$:

\begin{equation}  \label{1<k=n}
c(k) = 2 + 2^{n+1}.
\end{equation}

For $1 = k < n$:

\begin{equation}  \label{1=k<n}
c(1) = 2n(2^{1+2^{n-1}} - 3) + \sum_{r=2}^n \binom{r}{1} (-1)^{r-1} \binom{n%
}{r} 2^{r+1} (2^{2^{n-r}} - 1).
\end{equation}

For $1 = k = n$:

\begin{equation}
c(1) = 2+2\cdot 1(2^{1+2^{1-1}}-3)+0=4.  \label{1=k=n}
\end{equation}

\section{Randomly generating canalizing functions}

In simulating the behavior of random Boolean networks, it is important to be
able to randomly generate canalizing functions with a given bias $p$ \cite%
{ilyapost}. Our results in Section~4 allow us to do so by means of the
following algorithm:

Recall that for $1 \leq k \leq n$, $Pr_p(PCE_k)$ denotes the probability
that a randomly chosen Boolean function with bias $p$ is positively (but not
negatively) canalizing on exactly $k$ variables, and that $Pr_p(NCE_k)$
denotes the probability that a randomly chosen Boolean function with bias $p$
is negatively (but not positively) canalizing on exactly $k$ variables. Here
is the algorithm.

\vskip .1in

\noindent \textbf{Algorithm} \emph{CanalizingFunctionGenerator($p$)}

\vskip 0.1in

\begin{list}{ }{\itemsep 0.1in}
\item Let $q = 0$ with probability $\frac{Pr_p(BC)}{Pr_p(C)}$; 
for $1 \leq k \leq n$, let $q = k$ with probability $\frac{Pr_p(PCE_k)+ Pr_p(NCE_k)}{Pr_p(C)}$.

\item \textbf{if} $q \geq 1$ \textbf{then} let $r = 1$ with probability $\frac{Pr_p(PCE_k)}{Pr_p(PCE_k) + Pr_p(NCE_k)}$ and 
let $r = 0$ with probability $\frac{Pr_p(NCE_k)}{Pr_p(PCE_k) + Pr_p(NCE_k)}$.

\item \textbf{if} $q == 0$ \textbf{then}
\begin{list}{ }{\itemsep 0.1in}
\item Randomly pick an input variable $x_i$.
\item Randomly pick one of the two functions that are canalizing both ways on input $x_i$.
\item \textbf{return} the function $f$ that was just picked.
\end{list}

\item \textbf{else if} r == 1 \textbf{then}
\begin{list}{ }{\itemsep 0.1in} 
\item Randomly pick a subset $S$ of $[n]$ of size $q$.
\item Randomly pick a function $s: S \rightarrow \{0 , 1\}$.
\item For each input vector $x$ that contains some $x_i$ with $x_i = s(i)$ let $f(x) = 1$.
\item \textbf{repeat}
\begin{list}{ }{\itemsep 0.1in} 
\item For each of the remaining input vectors $x$ let independently and randomly $f(x) = 1$ with probability $p$ 
and let $f(x) = 0$ with probability $1-p$.
\end{list} 
\item \textbf{until} the resulting function $f$ is in $PCE_q$.
\item \textbf{return} $f$.
\end{list}
\item \textbf{else} // r == 0
\begin{list}{ }{\itemsep 0.1in} 
\item Randomly pick a subset $S$ of $[n]$ of size $q$.
\item Randomly pick a function $s: S \rightarrow \{0 , 1\}$.
\item For each input vector $x$ that contains some $x_i$ with $x_i = s(i)$ let $f(x) = 0$.
\item \textbf{repeat}
\begin{list}{ }{\itemsep 0.1in} 
\item For each of the remaining input vectors $x$ let independently and randomly $f(x) = 1$ with probability $p$ 
and let $f(x) = 0$ with probability $1-p$.
\end{list} 
\item \textbf{until} the resulting function $f$ is in $NCE_q$.
\item \textbf{return} $f$.
\end{list}

\end{list}

Note that the repeat $\dots$ until loops in this algorithm are necessary
since when parts of the vectors $x$ are assigned randomly, the resulting
function might, by chance, become canalizing for more than $q$ canalizing
variables. Should this occur, we would need to throw the function away and
generate another one.

\begin{table}[tbp] \centering%
\begin{tabular}{ll}
\hline
$n$ & $|C|$ \\ \hline
1 & $4$ \\ 
2 & $14$ \\ 
3 & $120$ \\ 
4 & $3514$ \\ 
5 & $1292\,276$ \\ 
6 & $103\,\allowbreak 071\,426\,294$ \\ 
7 & $\allowbreak 516\,\allowbreak 508\,833\,342\,\allowbreak 349\,371\,376$
\\ 
8 & $10\,889\,\allowbreak 035\,741\,470\,\allowbreak
030\,826\,695\,\allowbreak 916\,769\,153\,\allowbreak 787\,968\,498$ \\ 
9 & $\allowbreak 4.\,\allowbreak 168\,515\,213\times 10^{78}$ \\ 
10 & $5.\,\allowbreak 363\,123\,172\times 10^{155}$%
\end{tabular}%
\caption{The number of canalizing functions with $n$ input variables.}\label%
{TableKey}%
\end{table}%

\begin{thebibliography}{99}
\bibitem{wendt} P. Wendt, E. J. Coyle, N. Gallagher, \textquotedblleft Stack
Filters,\textquotedblright\ \textit{IEEE Trans. Acoust., Speech, Signal
Processing}, Vol. 34, pp. 898-911, 1986.

\bibitem{gabbouj} M. Gabbouj, P.-T. Yu, E. J. Coyle, \textquotedblleft
Convergence Behavior and Root Signal Set of Stack
Filters,\textquotedblright\ \textit{Circuit Systems and Signal Processing},
Vol. 11, No. 1, pp. 171-193, 1992.

\bibitem{yu} P-T. Yu, E. J. Coyle, \textquotedblleft Convergence Behavior
and N-Roots of Stack Filters,\textquotedblright\ \textit{IEEE Trans.
Acoust., Speech, Signal Processing,} Vol. 38, no. 9, 1990.

\bibitem{yucoyle} P-T. Yu, E. J. Coyle, \textquotedblleft The classification
and associative memory capability of stack filters,\textquotedblright\ 
\textit{IEEE Trans. Signal Processing,} Vol. 40, No. 10, pp. 2483-2497, 1992.

\bibitem{kauffmanbook} S. A. Kauffman, \textit{The origins of order:
Self-organization and selection in evolution}, Oxford University Press, New
York, 1993.

\bibitem{kauffman90} S. A. Kauffman, ``Requirements for Evolvability in
Complex Systems: Orderly Dynamics and Frozen Components,'' \textit{Physica D}%
, Vol. 42, pp. 135-152, 1990.

\bibitem{lynch} J. F. Lynch, ``On the Threshold of Chaos in Random Cellular
Automata,'' \textit{Random Structures and Algorithms}, Vol. 6, Nos. 2/3, pp.
239-260, 1995.

\bibitem{stauffer} D. Stauffer, \textquotedblleft On Forcing Functions in
Kauffman's Random Boolean Networks,\textquotedblright\ \textit{Journal of
Statistical Physics}, Vol. 46, Nos. 3/4, pp. 789-794, 1987.

\bibitem{ilyapost} I. Shmulevich, H. L\"{a}hdesm\"{a}ki, E. R. Dougherty, J.
Astola, W. Zhang, \textquotedblleft The role of certain Post classes in
Boolean network models of genetic networks,\textquotedblright\ \textit{%
Proceedings of the National Academy of Sciences of the USA}, Vol. 100, No.
19, pp. 10734-10739, 2003.

\bibitem{stern} M. D. Stern, ``Emergence of homeostasis and `noise
imprinting' in an evolution model,'' \textit{Proc. Natl. Acad. Sci. USA}
Vol. 96, pp. 10746-10751, 1999.

\bibitem{bornholdt} S. Bornholdt, K. Sneppen, \textquotedblleft Robustness
as an evolutionary principle,\textquotedblright\ \textit{Proc. Royal Soc.
London B}, Vol. 266, pp. 2281-2286, 2000.

\bibitem{glass} L. Glass, C. Hill, \textquotedblleft Ordered and disordered
dynamics in random networks,\textquotedblright\ \textit{Europhysics Letters,}
Vol. 41, pp. 599-604, 1998.

\bibitem{bhat} A. Bhattacharjya, S. Liang, \textquotedblleft Power-Law
Distributions in Some Random Boolean Networks,\textquotedblright\ \textit{%
Physical Review Letters} Vol. 77, pp. 1644-1647, 1996.

\bibitem{kauffmaninvestigations} S. A. Kauffman (2000) \textit{Investigations%
}, Oxford University Press, New York.

\bibitem{harris} S. E. Harris, B. K. Sawhill, A. Wuensche, S. Kauffman, 
\textit{Complexity, }Vol. 7, No. 4, pp. 23-40, 2002.

\bibitem{sawhill} B. K. Sawhill, S. A. Kauffman, \textit{Santa Fe Institute
Working Paper} 97-05-038, 1997.

\bibitem{harri} H. L\"{a}hdesm\"{a}ki, I. Shmulevich, O. Yli-Harja,
\textquotedblleft On Learning Gene Regulatory Networks Under the Boolean
Network Model,\textquotedblright\ \textit{Machine Learning}, Vol. 52, pp.
147-167, 2003.

\bibitem{spectral} I. Shmulevich, H. L\"{a}hdesm\"{a}ki, K. Egiazarian.
Spectral Methods for Testing Membership in Certain Post Classes and the
Class of Forcing Functions. \textit{IEEE Signal Processing Letters} (in
press).

\bibitem{aldana} M. Aldana, S. Coppersmith, L. P. Kadanoff,
\textquotedblleft Boolean Dynamics with Random Couplings,\textquotedblright\
in \textit{Perspectives and Problems in Nonlinear Science}, eds. Kaplan, E.,
Marsden, J. E., Sreenivasan, K. R. Springer, New York, pp. 23-89, 2002.

\bibitem{williamson} S. Gill Williamson. \textit{Combinatorics for Computer
Science.} Computer Science Press, Rockville, MD, 1985.
\end{thebibliography}
\end{document}